\documentclass[a4paper, headings=small, 10pt]{scrartcl}
\usepackage[english, british]{babel}
\usepackage[utf8]{inputenc}
\usepackage[T1]{fontenc}
\usepackage{newtxmath}
\usepackage{hyperref}
\usepackage{booktabs}
\usepackage{xcolor}
\usepackage[single]{accents}

\def\Z{{\mathbb{Z}}}

\def\fmax{\widetilde{f}}
\def\fmin{\undertilde{f}}

\let\phi=\varphi

\def\set#1{\{\, #1 \,\}}

\usepackage{listings}

\usepackage{listings}
\lstnewenvironment{code}
    {\lstset{}%
      \csname lst@SetFirstLabel\endcsname}
    {\csname lst@SaveFirstLabel\endcsname}
\title{Counting number-conserving cellular automata with radius 1}

\author{Markus Redeker\thanks{Hamburg, Germany. Email:
    \texttt{markus2.redeker@mail.de}}}
\begin{document}

\maketitle

\begin{abstract}\noindent
  This text is also a program that computes the number of
  one-dimensional number-conserving cellular automata with radius~1.
  At the end of the text, the numbers of such automata with up to~7
  cell states are shown.
\end{abstract}

\section{Introduction}

This is the description of a program that counts the one-dimensional
number-conserving cellular automata with radius 1, together with the
program source. The source files of this text\footnote{They can be
  downloaded from \url{https://arxiv.org/src/2605.31157}.} can be read
by a \LaTeX\ interpreter, which then generates the text you now read,
or by a Haskell compiler, which generates from it a program that
counts the cellular automata rules. The output of this program appears
at the end of this text, in Section~\ref{sec:results}.

In the text below, the Haskell code appears in the sections with a
gray backgrund, and the ordinary text around it explains it.

I will avoid here to explain the language Haskell (except for a few
details); for learning Haskell, the WikiBook \cite{Wikibooks2025} is a
better starting point.

\paragraph{Overview}

The rest of this text has three parts:
Section~\ref{sec:theory-counting} is an introduction into the theory
of number-conserving cellular automata of \cite{Redeker2022}, but only
as much as it is needed here. Section~\ref{sec:counting-rules} is a
description of the program and Section~\ref{sec:results} contains the
results.

\section{Theory of number-conserving cellular automata}
\label{sec:theory-counting}

\subsection{Cellular automata}

A one-dimensional cellular automaton is a discrete dynamic system that
at each point in time consists of a doubly infinite sequence of
objects, the \emph{cells}. Individual cells differ only by their
\emph{state}, which is an element of a finite set $\Sigma$.

The state of the whole dynomical system is then an element of the set
$\Sigma^\Z$ of \emph{configurations}. The content of a single configuration
$u \in \Sigma^\Z$ is written
\begin{equation}
  \label{eq:configuration}
  \dots u_{-2} u_{-1} u_0 u_1 u_2 u_3 \dots,
\end{equation}
with all the $u_i$ being elements of $\Sigma$. We will also use finite
sequences of cell states, which are written similarly.

The temporal evolution of such an automaton is given by a \emph{local
  rule} $\phi \colon \Sigma^{2r + 1} \to \Sigma$; the number $r$ is called the
\emph{radius} of the cellular automaton. The \emph{global rule}
$\hat{\phi}$, which transforms a configuration to another configuration,
is given by
\begin{equation}
  \label{eq:global-rule}
  \forall i \in \Z\colon \hat{\phi}(u)_i = \phi(u_{i-r}, \dots u_{i+r})\,.
\end{equation}

\subsection{Number conservation}

For number conservation, we define $\Sigma = \{ 0, \dots, C \}$ and call
$C$ the \emph{capacity} of the cellular automaton. Each cell is
thought as a container which may contain up to $C$ indistinguishable
particles. We distinguish between a cell state $\alpha \in \Sigma$ and its
\emph{particle content} $\# \alpha$, which is an element of $\Z$.

The particle content of a finite cell sequence
$a = \alpha_0 \dots \alpha_{k-1}$ is
$\# a = \# \alpha_0 + \dots + \# \alpha_{k-1}$, and with infinite cell sequences
as in~\eqref{eq:configuration} it is similar, but the particle content
may also be infinite.

We can then define a rule $\phi$ as \emph{number-conserving} if
\begin{equation}
  \label{eq:number-conserving}
  \# \hat{\phi}(u) = \# u
\end{equation}
for all configurations $u$ with finite particle content.

\subsection{Flow functions}

Now we can write down the central theorem on which relies the counting
algorithm for number-conserving cellular automata. Instead of the
radius of the cellular automata, it is expressed in terms of their
\emph{flow length} $\ell = 2r$ because with it, the
induction~\eqref{eq:free-halfflows} is less complicated.

\begin{quote}
  \textbf{Theorem.} To each number-conserving cellular automaton with
  rule $\phi$ belongs a unique \emph{flow function}
  $f \colon \Sigma^\ell \to \Sigma$ with $\ell = 2r$ as above, which in turn uniquely
  identifies $\phi$. \cite[Theorem~3.1]{Redeker2022}
\end{quote}

So we can count the flow functions instead of the transition rules,
which turns out to be easier. (The intuition behind the flow functions
is that when we think of the cells as containers with particles in
them, the flow function measures the number of particles at each time
step that cross a boundary between two cells. This number depends only
on an $\ell$-cell neighbourhood of the boundary. For more details see
\cite{Redeker2022} or \cite{Redeker2025}.)

\subsection{An induction over half-flows}

The central result that makes it possible to count the
number-conserving cellular automata is Theorem~4.3 of
\cite{Redeker2022}, which says that every flow function $f$ can be
constructed with the help of a sequence
\begin{equation}
  \label{eq:half-flow-sequence}
  (\fmin_0, \fmax_0), (\fmin_1, \fmax_1),
  \dots, (\fmin_\ell, \fmax_\ell) \,.
\end{equation}
of \emph{half-flow functions} $\fmin_i$,
$\fmax_i \colon \Sigma^i \to \Z$. The sequence ends with
$\fmin_\ell = \fmax_\ell = f$.

The pairs of half-flow functions $(\fmin_k, \fmax_k)$ must satisfy
several conditions. First,
\begin{subequations}
  \label{eq:free-halfflows}
  \begin{alignat}{3}
    \label{eq:lower-raising-recursion}
    \fmin_k(w_1 \dots w_k) &\leq \fmin_{k+1}(w)
    \leq \fmin_k(w_0 \dots w_{k-1}) + \# w_k, \\
    \label{eq:upper-raising-recursion}
    \fmax_k(w_0 \dots w_{k-1}) - (C - \# w_k) &\leq \fmax_{k+1}(w)
    \leq \fmax_k(w_1 \dots w_k) \\
    \intertext{must be true for $0 \leq k < \ell$ and $w = w_0 \dots w_k \in
      \Sigma^{k+1}$, and then,}
    \label{eq:fmin-bounds}
    0 &\leq \fmin_k(v) \leq \#v, \\
    \label{eq:fmax-bounds}
    0 &\leq \fmax_k(v) \leq \#v + (\ell - k)C, \\
    \label{eq:fmin-fmax-bounds}
    \fmin_k(v) &\leq \fmax_k(v) \leq \fmin_k(v) + (\ell - k)C
  \end{alignat}
\end{subequations}
must be true for $0 \leq k \leq \ell$ and $v \in \Sigma^k$. This way,
all flow functions can be constructed.

\subsection{Touch conditions}

But only with the conditions of~\eqref{eq:free-halfflows}, some of the
flow conditions will be created more than once. We must also have
\begin{subequations}
  \label{eq:half-flows-rec}
  \begin{align}
    \label{eq:lower-half-flow-rec}
    \fmin_k(v) &= \min \set{\fmin_{k+1}(\alpha v)\colon \alpha \in \Sigma}, \\
    \label{eq:upper-half-flow-rec}
    \fmax_k(v) &= \max \set{\fmax_{k+1}(\alpha v)\colon \alpha \in \Sigma},
  \end{align}
\end{subequations}
for $0 \leq k < \ell$ and $v \in \Sigma^k$ \cite[Equations~(16)]{Redeker2022}.
These conditions are almost redundant because, for example, the left
inequality in~\eqref{eq:lower-raising-recursion} is equivalent to
$\fmin_k(v) \leq \min \set{\fmin_{k+1}(\alpha v)\colon \alpha \in \Sigma}$. But to
strengthen this inequality to the equality that is required
in~\eqref{eq:half-flows-rec}, we must also have
\begin{subequations}
  \label{eq:touch-conditions}
  \begin{align}
    \label{eq:touch-conditions-min}
    \exists \alpha \in \Sigma \colon
    \fmin_k(v) &= \fmin_{k+1}(\alpha v), \\
    \label{eq:touch-conditions-max}
    \exists \alpha \in \Sigma \colon
    \fmax_k(v) &= \fmax_{k+1}(\alpha v)
  \end{align}
\end{subequations}
for all $v \in \Sigma^k$. Let us call these requirements the \emph{touch
  conditions} for $\fmin_k$ and $\fmax_k$.

\section{Counting the rules}
\label{sec:counting-rules}

Since we want to count the number of number-conserving cellular
automata with radius~2, we must set $\ell = 2$. The program is written
especially for this case and cannot easily be changed for other values
of $\ell$.

\subsection{Preliminaries}

The program begins with some \lstinline{import} statements.

\begin{code}
import System.Environment
import Data.Array.Unboxed
\end{code}

\subsection{The \lstinline{main} function}

The program starts with the function \lstinline{main}. This function
reads the number of states from the command line and prints the number
of number-conserving rules with radius~1 and the required number of
states to the standard output. The actual computation is done by the
function \lstinline{count0}.

\begin{code}
main :: IO ()
main =
  do args <- System.Environment.getArgs
     let states = read (args !! 0)
     let numRules = count0 (states - 1)
     putStrLn (show numRules)
\end{code}

\subsection{Construction of the half-flows $\protect\fmin_0$ and
  $\protect\fmax_0$}

The program counts the number-conserving cellular automata in three
nested loops: The outermost loop is realised in the function
\lstinline{count0}, which finds all permissible choices for $\fmin_0$
and $\fmax_0$. For each of these choices, it computes how many
construction sequences~\eqref{eq:half-flow-sequence} start with 
$(\fmin_0, \fmax_0)$. Then it returns the sum of these numbers.

The half-flows $\fmin_0$ and $\fmax_0$ are functions that take an
element of $\Sigma^0$ as their parameter, which means that they are in fact
constanats. For them, the conditions of~\eqref{eq:free-halfflows}
reduce considerably to
\begin{align}
  \label{eq:halfflows-0}
  \fmin_0 &= 0, & 0 \leq &\fmax_0 \leq 2 C\,.
\end{align}
The function \lstinline{count0}, which refers to the function
\lstinline{count1} below for the inner counting loops, is therefore
quite simple:
\begin{code}
count0 :: Int -> Integer
count0 capacity =
  sum [count1 capacity fmax0 | fmax0 <- [0..2 * capacity]]
\end{code}
Since the number cellular automata rules becomes very large even for
small cell capacities, the function \lstinline{count0} and later
functions return a value of type \lstinline{Integer}, which is
Haskell's data type for integers of unrestricted size.

\subsection{Construction of the half-flows $\protect\fmin_1$ and
  $\protect\fmax_1$}

The functions $\fmin_1$ and $\fmax_1 \colon \Sigma \to \Sigma$ are represented in
the program as one-dimensional arrays of length $C + 1$; we use the
data type \lstinline{Flow1} for them. The type alias \lstinline{State}
is always used when a cell state is represented.

\begin{code}
type State = Int
type Flow1 = UArray Int State
\end{code}

Now the equations of~\eqref{eq:free-halfflows} become
\begin{subequations}
  \label{eq:halfflows1}
  \begin{alignat}{3}
    \label{eq:lower-raising-recursion1}
    \bigl(
    \fmin_0 &\leq \fmin_1(\alpha)
              \leq \fmin_0 + \# \alpha,
              \bigr) \\
    \label{eq:upper-raising-recursion1}
    \fmax_0 - (C - \# \alpha) &\leq \fmax_1(\alpha)
    \leq \fmax_0, \\
    \label{eq:fmin-bounds1}
    0 &\leq \fmin_1(\alpha) \leq \#\alpha, \\
    \label{eq:fmax-bounds1}
    0 &\leq \fmax_1(\alpha) \leq \#\alpha + (\ell - k)C, \\
    \label{eq:fmin-fmax-bounds1}
    \fmin_1(\alpha) &\leq \fmax_1(\alpha) \leq \fmin_1(\alpha) + (\ell - k)C\,.
  \end{alignat}
\end{subequations}
Here I have put \eqref{eq:lower-raising-recursion1} in braces because
it is redundant to~\eqref{eq:fmin-bounds1}; this is so because we know
from~\eqref{eq:halfflows-0} that $\fmin_0 = 0$.

The remaining inequalities are then incorporated into the function
\lstinline{valueBoundsFor}, where
\lstinline{valueBoundsFor}~$C\ \alpha\ \fmax_0$ creates a list of all the
possible choices for the pairs $(\fmin_1(\alpha), \fmax_1(\alpha))$ when the
value of $\fmax_0$ is already given.
\begin{code}
valueBoundsFor :: Int -> State -> State -> [(State, State)]
valueBoundsFor capacity alpha fmax0 =
  let fmin1_max = minimum [alpha,
                           fmax0]
  in
    [(fmin1, fmax1) |
     fmin1 <- [0..fmin1_max],
     let fmax1_min = maximum [fmax0 - (capacity - alpha),
                              fmin1],
     let fmax1_max = minimum [fmax0,
                              fmin1 + capacity],
     fmax1 <- [fmax1_min..fmax1_max]]
\end{code}
We do not need to take care of the touch conditions since they are
automatically satisfied: To make \eqref{eq:touch-conditions-min} true,
we must find an $\alpha$ with $\fmin_0 = \fmin_1(\alpha)$, and we see
from~\eqref{eq:lower-raising-recursion1} that the $\alpha$ with
$\# \alpha = 0$ is the right choice. For~\eqref{eq:touch-conditions-max},
we must find a solution for $\fmax_0 = \fmax_1(\alpha)$, and
\eqref{eq:fmin-bounds1} tells us that the $\alpha$ with
$\# \alpha = C$ is the right choice.

For the second loop, we therefore need a function that constructs all
functions $\fmin_1$ and $\fmax_1$ according to~\eqref{eq:halfflows1}
and then for each pair $(\fmin_1, \fmax_1)$ counts the number of flow
functions that belong to it. This does the function
\lstinline{count1}. It creates with the help of the function
\lstinline{valueBoundsFor} all possible pairs
$(\fmin_1(\alpha), \fmax_1(\alpha))$ and then, in a complicated process, it
transforms them into a list of all possible pairs of functions
$(\fmin_1, \fmax_1)$. The function \lstinline{count2} then determines
the number of flow functions that belong to such a pair. And at the
end, all these numbers are summed up.

\begin{code}
count1 :: Int -> State -> Integer
count1 capacity fmax0 =
  let valueBounds = [valueBoundsFor capacity alpha fmax0 |
                     alpha <- [0..capacity]]
      toHalfFlow a = listArray (0, capacity) a
      toHalfFlows (a, b) = (toHalfFlow a, toHalfFlow b)
  in
    sum [count2 capacity fmin1 fmax1 |
         (fmin1, fmax1) <- map (toHalfFlows . unzip) $ sequence valueBounds]
\end{code}

\subsection{Counting the flow functions}

The task of the innermost loop would be the construction of all pairs
$(\fmin_2, \fmax_2)$. But we know already that $\fmin_2 = \fmax_2 = f$
for every sequence~\eqref{eq:half-flow-sequence}, so we need to find
only all possible $f$ functions. And this time, we do not construct
but only count them.

The inequalities~\eqref{eq:free-halfflows} now become
\begin{subequations}
  \label{eq:halfflows2}
  \begin{alignat}{3}
    \label{eq:lower-raising-recursion2}
    \fmin_1(\beta) &\leq f(\alpha \beta) \leq \fmin_1(\alpha) + \# \beta, \\
    \label{eq:upper-raising-recursion2}
    \fmax_1(\alpha) - (C - \# \beta) &\leq f(\alpha \beta) \leq \fmax_1(\beta), \\
    \label{eq:fmin-bounds2}
    0 &\leq f(\alpha \beta) \leq \#\alpha + \# \beta\,.
  \end{alignat}
\end{subequations}
In them, \eqref{eq:fmin-bounds2} is the translation of
both~\eqref{eq:fmin-bounds} and~\eqref{eq:fmax-bounds}, because here
they have the same meaning. And~\eqref{eq:fmin-fmax-bounds} here means
that $\fmin_2 = \fmax_2$, which we know already.

But the touch conditions are no longer automatically satisfied. For
$k = 1$, they take the form
\begin{subequations}
  \label{eq:touch-conditions1}
  \begin{align}
    \label{eq:touch-conditions-min1}
    \exists \alpha \in \Sigma \colon
    \fmin_1(\beta) &= f(\alpha \beta), \\
    \label{eq:touch-conditions-max1}
    \exists \alpha \in \Sigma \colon
    \fmax_1(\beta) &= f(\alpha \beta)\
  \end{align}
\end{subequations}
for all $\beta \in \Sigma$. To take them into account, we create a parametrized
version of~\eqref{eq:halfflows2}, namely
\begin{subequations}
  \label{eq:halfflows2par}
  \begin{alignat}{3}
    \label{eq:lower-raising-recursion2par}
    \fmin_1(\beta) + i &\leq f(\alpha \beta)
                     \leq \fmin_1(\alpha) + \# \beta, \\
    \label{eq:upper-raising-recursion2par}
    \fmax_1(\alpha) - (C - \# \beta) &\leq f(\alpha \beta)
                              \leq \fmax_1(\beta) - j, \\
    \label{eq:fmin-bounds2par}
    0 &\leq f(\alpha \beta) \leq \#\alpha + \# \beta\,.
  \end{alignat}
\end{subequations}
It is implemented in the function \lstinline{numValuesFor} below. This
function computes the smallest and the largest possible value for
$f(\alpha \beta)$ according to~\eqref{eq:lower-raising-recursion2par} and then
computes from them the number of different values that $f(\alpha \beta)$ can
take under the given conditions.

\begin{code}
numValuesFor :: Int -> Flow1 -> Flow1
  -> State -> State -> Int -> Int -> Integer
numValuesFor capacity fmin1 fmax1 alpha beta i j =
  let val_min = maximum [fmin1 ! beta + i,
                         fmax1 ! alpha - (capacity - beta),
                         0]
      val_max = minimum [fmin1 ! alpha + beta,
                         fmax1 ! beta - j,
                         alpha + beta]
  in
    toInteger $ max 0 (val_max + 1 - val_min)
\end{code}

Next we let $\nu_{ij}(\alpha
\beta)$ be the number of choices for $f(\alpha
\beta)$ that satisfy the inequalities of~(\ref{eq:halfflows2par}) or in
other words, the value of \lstinline{numValuesFor} $C$
$\fmin_1$ $\fmax_1$ $\alpha$ $\beta$ $i$ $j$. Then the expression
\begin{equation}
  \label{eq:nu'}
  \nu'_{ij}(\beta) = \prod_{\alpha\in\Sigma} \nu_{ij}(\alpha \beta)
\end{equation}
is the number of choices for a function $\alpha \mapsto f(\alpha
\beta)$ that satisfies~\eqref{eq:halfflows2par}. More specifically,
$\nu'_{00}(\beta)$ is the number of such functions that
satisfy~\eqref{eq:halfflows2},
$\nu'_{10}(\beta)$ the number of functions where also $\fmin_1(\beta) \neq f(\alpha
\beta)$ for all $\alpha$, and so on. Therefore, $\nu'_{00}(\beta) -
\nu'_{10}(\beta)$ counts all functions that also satisfy the touch
condition~(\ref{eq:touch-conditions-min1}), and, by the law of
inclusion and exclusion,
\begin{equation}
  \label{eq:n-sum}
  \nu''(\beta) = \nu'_{00}(\beta) - \nu'_{01}(\beta) - \nu'_{10}(\beta) + \nu'_{11}(\beta)
\end{equation}
is the number of functions $\alpha \mapsto f(\alpha
\beta)$ that satisfy~\eqref{eq:halfflows2} and the touch conditions. And
the number of flow functions
$f$ that satisfy both families of conditions is obviously $\prod_{\beta\in\Sigma}
\nu''(\beta)$.

The function \lstinline{count2} below then computesn this product.
\begin{code}
count2 :: Int -> Flow1 -> Flow1 -> Integer
count2 capacity fmin1 fmax1 =
  let num beta i j =
        product [numValuesFor capacity fmin1 fmax1 alpha beta i j |
                 alpha <- [0..capacity]]
  in
    product [(num beta 1 1) - (num beta 1 0)
              - (num beta 0 1) + (num beta 0 0) | beta <- [0..capacity]]
\end{code}

\section{Results}
\label{sec:results}

The results of these computations are shown in the following table.

\begin{center}
  \begin{tabular}{rr}
    \toprule
    States & Number of Rules \\
    \midrule
    1 & 1                        \\
    2 & 5                        \\
    3 & 144                      \\
    4 & 89588                    \\ 
    5 & 1876088314               \\ 
    6 & 2127241618152346         \\ 
    7 & 177778540782808970125334 \\ 
    \bottomrule
  \end{tabular}
\end{center}

The only values in this table that are new are those for 6 and 7
states; the rest occurs already in \cite{Wolnik2025}.\footnote{The
  value for 4 states in \cite{Wolnik2025}, which differs from that
  given here, is actually an error that occurred when copying it from
  the original source. (B. Wolnik, pers.~comm.)}


\bibliography{webresources,../references}

\begin{thebibliography}{1}

\bibitem{Redeker2022}
Markus Redeker.
\newblock Number conservation via particle flow in one-dimensional cellular
  automata.
\newblock {\em Fundamenta Informaticae}, 187(1):31--59, 2022.

\bibitem{Redeker2025}
Markus Redeker.
\newblock An invitation to number-conserving cellular automata.
\newblock In Andrew Adamatzky, Georgios~Ch. Sirakoulis, and Genaro~J. Martinez,
  editors, {\em Advances in Cellular Automata: Volume 1: Theory}, pages
  457--472. Springer Nature Switzerland, Cham, 2025.

\bibitem{Wikibooks2025}
Wikibooks.
\newblock Haskell.
\newblock \url{https://en.wikibooks.org/wiki/Haskell}, accessed 2026-06-17,
  2025.

\bibitem{Wolnik2025}
Barbara Wolnik, Witold Bo{\l}t, and Bernard~De Baets.
\newblock Recent insights into number-conserving cellular automata.
\newblock In Andrew Adamatzky, Georgios~Ch. Sirakoulis, and Genaro~J.
  Mart\'{\i}nez, editors, {\em Advances in Cellular Automata: Volume 1:
  Theory}, pages 399--422. Springer Nature Switzerland, 2025.

\end{thebibliography}
\end{document}